\documentclass[a4paper,12pt]{article}
\usepackage{graphicx}
\usepackage{mathrsfs}
\usepackage{tabularray}
\usepackage{indentfirst}
\usepackage[utf8]{inputenc}
\usepackage[english,main=russian]{babel}
\usepackage{amsmath,amssymb}
\usepackage{dsfont}
\usepackage{subfigure}
\usepackage{cite}
\usepackage{booktabs}
\usepackage[unicode,linktocpage=true,plainpages=false,pdfpagelabels=false]{hyperref}
\usepackage{todonotes}
\usepackage{multirow}

\title{Динамическое описание фазового перехода в сверхпроводящее состояние}
\author{
Л.~А.~Гостева\textsuperscript{1,}\thanks{E-mail: lagosteva1999@gmail.com} \and 
М.~Ю.~Налимов\textsuperscript{1,2} \and
А.~С.~Яшугин\textsuperscript{1}
}
\date{
$^1$\it\small Санкт-Петербургский Государственный Университет, Университетская наб., 7/9, Санкт-Петербург, 199034, Россия\\
$^2$Лаборатория теоретической физики им. Н.Н. Боголюбова ОИЯИ, ул. Жолио-Кюри, 6, Дубна, Московская обл., 141980, Россия
}

\begin{document}

\maketitle

\begin{abstract}
В статье приводятся динамические уравнения, справедливые в окрестности фазового перехода в сверхпроводящее состояние. При написании уравнений были учтены возможные влияния полей магнитного взаимодействия носителей заряда и температурных флуктуаций. Обсуждается тип исследуемого фазового перехода. Приводятся соображения в пользу применимости стохастической динамической модели~A для описания динамики данного фазового перехода.

\textbf{Ключевые слова:} сверхпроводимость, критические явления, критическая динамика, фазовый переход, микроскопическая модель, ренормализационная группа, $4-\varepsilon$ разложение, MS схема ренормировки.
\end{abstract}

\section{Введение}
Целью данной работы является создание и анализ феноменологических стохастических уравнений для описания динамики перехода в сверхпроводящее состояние. Как правило, для описания динамики в сверхпроводниках использовались уравнения Гросса--Питаевского (ГП) или зависящие от времени уравнения Гинзбурга--Ландау (ГЛ) \cite{kopnin, shanenko}.

Уравнение ГП, изначально созданное для описания сверхтекучести в бозонных системах, по сути, является нелинейным уравнением Шредингера для волновой функции $\psi(\mathbf{x},t)$, описывающей конденсат <<составных бозонов>> \cite{kopnin}:
\begin{equation}
    i \partial_t \psi = 
    - \left( \frac{\Delta}{2m_B} + \mu_B \right) \psi + U_0 |\psi^2| \psi ,
\label{GP}
\end{equation}
где $\hbar \equiv 1$, $m_B$ --- масса бозона, $\mu_B$ --- химический потенциал бозона, $U_0>0$ --- константа, отражающая парное взаимодействие бозонов в локальном приближении: $U(\mathbf{x} - \mathbf{x}^\prime) = U_0 \delta (\mathbf{x} - \mathbf{x}^\prime)$. Возникает вопрос о смысле входящих в уравнение параметров. Аккуратный вывод уравнения ГП из уравнений Боголюбова--де Жена для фермионной системы \cite{pieri2003} показывает связь фермионных массы $m_F$ и химпотенциала $\mu_F$ с параметрами «составных бозонов» (<<куперовских пар>>): $m_B=2m_F$, $\mu_B=2\mu_F$. Даже такой вывод нельзя назвать полностью микроскопическим обоснованием, поскольку в нем используется самосогласование с выбором анзаца для параметра щели. Заметим, что уравнение ГП обосновано лишь при околонулевой температуре \cite{shanenko, pieri2003}. Действительно, ГП не учитывает возбуждения бозонов, которые возникли бы при конечной температуре. Это, в том числе, проявляется в Галилей-инвариантности ГП, что ведет к отсутствию в такой модели максимальной скорости, при превышении которой явление сверхпроводимости должно разрушаться из-за возбуждений \cite{kopnin}. В работе \cite{hs} для случая $T=0$ также был проделан переход от временных нелинейных уравнений Шредингера для фермионных полей к уравнению ГП для бозонных полей в пределе сильной связи между фермионами. Но главный недостаток ГП в контексте рассматриваемой задачи заключается в том, что оно годится лишь для описания динамических квантовомеханических эффектов и вообще непригодно для описания критической динамики, в которой главным эффектом является аномально большие в критической области времена релаксации.

Теория сверхпроводимости Гинзбурга--Ландау, напротив, была создана для описания явлений вблизи критической точки, однако, будучи теорией среднего поля, работает только вне критической области. Простейшая модель релаксационной динамики с учетом калибровочной инвариантности дает временное уравнение ГЛ \cite{kopnin}:
\begin{equation}
    -\Gamma \left( \partial_t - 2ie_F\phi \right) \psi = 
    \alpha\psi + \beta|\psi^2|\psi - \gamma \left( \nabla + 2ie_F\mathbf{A} \right)^2 \psi ,
\label{GL}
\end{equation}
где $\hbar, c \equiv 1$, $\psi$ также имеет смысл некой волновой функции конденсата <<составных бозонов>>, нормированной на энергетическую щель, $\phi$ и $\mathbf{A}$ --- скалярный и векторный потенциалы, $e_F$ --- заряд фермиона, $\Gamma$, $\beta$ и $\gamma$ --- положительные константы, $\alpha$ пропорциональна отклонению температуры от критической. Главное отличие (\ref{GL}) от (\ref{GP}) заключается в вещественности коэффициента перед производной по времени. Но, как известно \cite{Vasiljev0}, приближение среднего поля недостаточно для описания критических явлений (в частности, получаемые критические индексы далеки от истинных). Таким образом, представляется затруднительным извлечь из уравнений ГП или ГЛ динамические характеристики, связанные именно с переходной областью.

В дальнейшем критическая динамика сверхпроводящего фазового перехода исследовалась \cite{Dudka_gauge_fields} на основании модели многокомпонентной скалярной электродинамики, выбранной из феноменологических соображений. Однако в данных работах анализ свелся главным образом к изучению критической статики, ибо возникла проблема определения самого типа фазового перехода. Так, в работах \cite{Antonov_2016, Dudka_gauge_fields} в однопетлевом приближении было показано, что наличие калибровочного поля ${\bf A}$ приводит к изменению характера фазового перехода, который становится переходом I рода.  

Таким образом, вопрос о роде фазового перехода до сих пор следует считать открытым \cite{Kleinert05}. Модели типа БКШ приводят к непрерывному фазовому переходу, учет магнитного взаимодействия приводит к переходу первого рода, как и учет возможных слоев и подрешеток в системе. В последнем случае род данного фазового перехода подтвержден пятипетлевыми вычислениями \cite{Kalagov}, но в случае учета магнитного взаимодействия между носителями заряда ситуация не столь определенна. В работе \cite{Folk1996} утверждается, что учет двухпетлевых поправок в модели типа скалярной электродинамики возвращает нас к фазовому переходу второго рода. Однако в цитируемой работе применен совершенно нетрадиционный, непертурбативный метод учета старших вкладов теории возмущений, к выводам данной работы надо относиться осторожно. 

Экспериментальные работы для динамических характеристик рассматриваемого фазового перехода дают значения динамического индекса $z$ от 1.5 до 2.3 (см. \cite{Dudka_gauge_fields}), указывая на существование фазового перехода второго рода. Примерно такой же разброс дают и компьютерные эксперименты, основанные на методе Монте-Карло.

В работе \cite{Komarova_2013} был строго обоснован переход от микроскопического описания фермионной системы на языке грассмановых переменных к описанию в терминах бозонных полей, которые являются безмассовыми модами в критической точке. Показано, что среднее значение этих полей действительно является параметром порядка фазового перехода --- аномальным значением среднего пары фермионных полей. Соответственно, именно в таких переменных имеет смысл описывать динамику фазового перехода. В данной работе мы строим стохастические уравнения, основываясь на предложенном в \cite{Komarova_2013} действии, параметры которого имеют явную связь с микроскопическими параметрами \cite{Kalagov}. 

Другой важной стороной динамического описания является учет дополнительных мягких мод, которые были бы существенны при рассмотрении фазового перехода. Наличие других полей может сильно изменить картину фазового перехода. Постановка же эксперимента, на результат которого можно было бы опираться в поисках корректной модели, крайне сложна в виду необходимости рассмотрения чрезвычайно малой окрестности критической точки $|T-T_c| \sim 10^{-6}K$\cite{Halperin_gauge_fields}. 

В данной работе на основе статической модели мы построим уравнения динамики параметра порядка с учетом теплопроводности и магнитных взаимодействий. Для этого включим в статическую модель \cite{Komarova_2013} соответствующие поля. Такая расширенная модель не рассматривалась ранее и потребует ренормгруппового анализа, который мы проведем в однопетлевом приближении.

Структура статьи следующая. В разделе 2 мы обсудим статику: дадим основные идеи перехода от действия, описывающего систему взаимодействующих фермионов, к действию, которое описывает критическую окрестность и содержит параметр порядка в качестве переменной. Также мы обсудим, как включить в это действие дополнительные поля для учета флуктуаций температуры и учета магнитных взаимодействий. В разделе 3 мы проведем ренормгрупповое исследование получившейся модели для определения возможных критических режимов и типа фазового перехода. Наконец, в разделе 4 мы перейдем к динамике и на основе полученного статического действия выпишем уравнения динамики упомянутых полей, и обсудим частные случаи, в которых динамика является критической (то есть имеет место фазовый переход второго рода).

\section {Статическое действие в терминах поля параметра порядка}

Идея введения бозонных полей и получения эффективного действия, описанного в \cite{Komarova_2013}, заключается в следующем: в формализме температурных функций Грина рассматривается статическое действие квантово-полевой модели с 4-фермионным взаимодействием 
\begin{equation}
    S = \psi^+_{\alpha} \left( \partial_t - \frac{\Delta}{2m} - \mu \right) \psi_{\alpha} - 
    \frac{\lambda}{2}\left(\psi^+_{\alpha} \psi_{\alpha}\right) \left(\psi^+_{\beta} \psi_{\beta}\right),
\label{spin_action}
\end{equation}
где $\psi_{\alpha}\left( \mathbf{x}, t \right)$, $\psi^+_{\alpha}\left( \mathbf{x}, t \right)$ --- комплексно-сопряженные фермионные поля, которые являются грассмановыми переменными, $\alpha=1,...,r$ --- индекс, связанный со спинами, подрешетками и/или слоистыми структурами материала, $\mathbf{x}$ --- координата, $t$ --- получаемое в результате евклидового разворота <<мнимое время>>, меняющееся в интервале $[0,\beta]$, где $\beta=1/(kT)$, $\mu$ --- химический потенциал, $\lambda$ --- константа четырехфермионного взаимодействия, являющегося следствием электрон-фононного взаимодействия или иного, вызывающего притяжение между электронами. Здесь и далее необходимые интегрирования по $t$ и $\mathbf{x}$ подразумеваются. Заметим, что локальность взаимодействия является совершенно естественной в предлагаемом подходе: мы всюду ориентируемся на описание критического поведения или используем гидродинамическое приближение, и в обоих случаях во всевозможных нелокальных характеристиках следует учитывать лишь главные члены по волновым векторам (импульсам). Преобразованием Хаббарда--Стратоновича в действие (\ref{spin_action}) вводятся новые бозонные поля $\chi_{\alpha\beta}$ и $\chi_{\alpha\beta}^+$ --- комплексные антисимметричные матрицы ранга $r$:
\begin{equation} 
    S = \psi^+_{\alpha} \left( \partial_t -\frac{\Delta}{2m}-\mu \right) \psi_{\alpha} + 
    \chi_{\beta\alpha}^+\chi_{\alpha\beta} +
    \sqrt{\frac{\lambda}{2}}\chi_{\alpha\beta}\left( \psi^+_{\alpha} \psi^+_{\beta} \right) +
    \sqrt{\frac{\lambda}{2}}\chi_{\alpha\beta}^+\left( \psi_{\alpha} \psi_{\beta}\right),
\label{action_spin_chi}
\end{equation}
Уравнения Швингера показывают, что средние значения введенных полей непосредственно связаны с параметром порядка сверхпроводящего фазового перехода \cite{Komarova_2013}:
\begin{gather}
    \left\langle \chi_{\alpha\beta}^+ \right\rangle =
    \sqrt{\frac{\lambda}{2}} \left\langle \psi_{\alpha}^+ \psi_{\beta}^+ \right\rangle, \ 
    \left\langle \chi_{\alpha\beta} \right\rangle = 
    \sqrt{\frac{\lambda}{2}} \left\langle \psi_{\alpha} \psi_{\beta} \right\rangle,
\end{gather}
поэтому данные поля действительно можно рассматривать как микроскопические аналоги параметра порядка, а модель (\ref{action_spin_chi}) является микроскопической моделью с явно выделенными критическими модами $\chi$ и $\chi^+$.
Для получения эффективного действия для критических мод выполняют гауссово интегрирование по полям $\psi$, $\psi^+$ и отбирают ИК-существенные члены. В результате получается эффективное статическое действие квантово-полевой модели, описывающее поведение системы в окрестности точки фазового перехода \cite{Komarova_2013}: 
\begin{gather}    
    S \left( \chi^+, \chi \right) =
    c_1 \mathrm{Tr}\left(\nabla \chi^{+}\nabla \chi\right) +
    \tau \mathrm{Tr}\left(\chi^{+}\chi\right) +
    \frac{{\tilde g}_{1}}{4}(\mathrm{Tr}(\chi\chi^{+}))^2 + 
    \frac{g_{2}}{4}\mathrm{Tr}\left(\chi\chi^{+}\chi\chi^{+}\right),
\label{action_0}
\end{gather}
здесь ${\tilde g}_{1}$ и $g_{2}$ --- константы связи (заряды), $\tau \sim |T-T_c|$ --- отклонение температуры от критического значения $T_c$. Член с зарядом ${\tilde g}_{1}$ исходно отсутствует и вводится для мультипликативной ренормируемости модели (он генерируeтся в результате ренормировки). В дальнейшем константа $c_1$ в нашей работе устранена растяжением полей. Подчеркнем, что все введенные константы $c_1$, $g_{2}$, $\tau$ вычислены через параметры исходной модели (\ref{spin_action}) как функции физических параметров: температуры, химического потенциала, параметра электронного взаимодействия \cite{Kalagov}. В работе \cite{Komarova_2013} показывается, что поля $\chi,\chi^+$ оказываются безмассовыми критическими модами в точке фазового перехода, и, следовательно, в их терминах корректно описывается фазовый переход. В этот момент мы отказываемся в описании от формализма температурных функций Грина. Действительно, в критической или околокритической области можно пренебречь вкладами всех мацубаровских частот за исключением нулевой, тем самым поля $\chi^+, \chi$ можно считать независящими от времени, и действие теории становится статическим. Именно это действие, а не получаемое, по сути, на основании симметрийных соображений (как, например, в \cite{Dudka_gauge_fields}), будет положено в основу нашей статьи. 

Отметим, что сама идея перехода от фермионных полей к бозонным, с последующим разложением по импульсам, не нова (см. \cite{Popov, LarkinVarlamov}), однако в предыдущих работах этим и ограничивались. Такой подход в духе теории среднего поля не справедлив для критической области, что особенно важно для описания высокотемпературных сверхпроводников, где эта область широка. Процедура ренормировки же позволяет описывать критическую область, что и является нашей целью.

Обратим также внимание на то, что переменные $\chi^+, \chi$ связаны с парами фермионных переменных лишь в среднем. Для того, чтобы ввести переменные $\kappa^+, \kappa$, точно равные таким парам, можно воспользоваться следующим тождеством:
\begin{equation} 
    e^{-S(\psi^+,\psi)} = 
    \int D\kappa^+ D\kappa D\chi^+ D\chi 
    e^{-S(\psi^+,\psi) + (\kappa - \psi\psi)\chi^+ + (\kappa^+ - \psi^+\psi^+)\chi}.
\end{equation}
Здесь мы ввели функциональные дельта-функции, обеспечивающие равенства $\kappa^+ = \psi^+ \psi^+$, $\kappa = \psi \psi$, и представили их в виде аналога преобразования Фурье, вводя вспомогательные поля $\chi^+, \chi$. Интегрирование по всем полям кроме $\kappa^+, \kappa$ и выбор лишь существенных слагаемых приведет нас к действию $S\left(\kappa^+, \kappa \right)$ вида (\ref{action_0}), но с отличными значениями параметров. Далее мы будем работать с действием $S \left( \chi^+, \chi \right)$, но анализ действия $S \left( \kappa^+, \kappa \right)$ возможно провести аналогичным образом, при этом получаемые характеристики фазового перехода, которыми мы интересуемся, будут, очевидно, такими же.

Наличие спина у фермионов приводит к магнитному взаимодействию между частицами (электростатическое взаимодействие традиционно считается несущественным и учитывается разве что в качестве поправок к химическому потенциалу). Вследствие калибровочной инвариантности включение магнитного взаимодействия в действие согласно работам \cite{Halperin_gauge_fields}, \cite{Antonov_2016} осуществляется простой заменой пространственных производных на ковариантные (что вполне согласуется с описанным выше выводом (\ref{action_0})). В результате мы получаем действие: 
\begin{gather} 
S\left( \chi,\chi^+,\mathbf{A} \right) = 
\mathrm{Tr}\left(\left(\nabla+ie\mathbf{A}\right)\chi^{+}\left(\nabla-ie\mathbf{A}\right)\chi\right) +
\tau \mathrm{Tr}\left(\chi^{+}\chi\right) +
\frac{g_{1}}{4}\left(\mathrm{Tr}\left(\chi\chi^{+}\right)\right)^2 + \nonumber\\
+\ \frac{g_{2}}{4}\mathrm{Tr}\left(\chi\chi^{+}\chi\chi^{+}\right) +\
\frac{1}{2}\left(\nabla \times  \mathbf{A}\right)^2. 
\label{action_a}
\end{gather}
Здесь используется калибровка Кулона ($\nabla \cdot \mathbf{A} = 0$), $\mathbf{A}$ --- векторный потенциал, $e$ --- заряд электрона. Однако в \cite{Halperin_gauge_fields, Antonov_2016} в результате однопетлевых вычислений было показано, что в данной или ей эквивалентной модели отсутствует инфракрасно устойчивая фиксированная точка, что может трактоваться как наличие фазового перехода первого рода, вместо второго. Поэтому возникает вопрос: как повлияют на тип фазового перехода иные гидродинамические переменные, в первую очередь поле флуктуаций температуры.

Введение поля температуры произведем стандартным образом, аналогично построению динамической модели C в классификации \cite{hohenberg-1977}. С помощью преобразования Хаббарда--Стратоновича включим в набор рассматриваемых полей дополнительное поле $m$, и перепишем статическое действие (\ref{action_a}) в виде: 
\begin{gather} 
S\left(\chi,\chi^+,\mathbf{A},m \right)=\mathrm{Tr}\left(\left(\nabla+ie\mathbf{A}\right)\chi^{+}\left(\nabla-ie\mathbf{A}\right)\chi\right)+\tau \mathrm{Tr}\left(\chi^{+}\chi\right)+\frac{g_{1}}{4}\left(\mathrm{Tr}\left(\chi\chi^{+}\right)\right)^2 + \nonumber\\
+\ \frac{g_{2}}{4}\mathrm{Tr}\left(\chi\chi^{+}\chi\chi^{+}\right) 
+\frac{1}{2}\left(\nabla \times  \mathbf{A}\right)^2+g_{3} \mathrm{Tr}\left( \chi\chi^{+} \right)m+\frac{m^2}{2},
\label{action_m_a}
\end{gather}
где $g_1=\tilde g_1 - 2 g_3^2$. Условие на положительную определенность взаимодействия в действии (\ref{action_m_a}) определяет область устойчивости модели:
\begin{equation}
    g_2 + rg_1 >0, \ e^2>0, \ g_3>0.
\label{stability}
\end{equation} Из явного вида действия (\ref{action_m_a}) очевидно, что связанный с отклонением температуры от критической параметр $\tau$ входит в сумме с полем $m$ и может быть включен в него. Это показывает, что поле $m$ действительно связано с полем флуктуаций температуры. Однако более детальный феноменологический анализ вывода модели C \cite{hohenberg-1977} приводит к заключению, что поле $m$ представляет собой локальное поле некоторого сохраняющегося термодинамического потенциала, связанного с флуктуациями температуры термодинамическими соотношениями. Коэффициент пропорциональности может быть рассчитан из микроскопической модели, но в задачу данной статьи это не входит.

\section{Ренормгрупповой анализ статического действия}

Мы собираемся описывать критическую динамику равновесных флуктуаций в окрестности точки фазового перехода в сверхпроводящее состояние, поэтому сперва следует исследовать полученную статическую модель (\ref{action_m_a}) для определения существующих в ней критических режимов: действительно ли в данной модели проявляется фазовый переход 2 рода? Ограничимся случаем обычных носителей заряда --- электронов в однородной среде. Этому соответствует значение $r=2$ (фермионы со спином $1/2$). Тогда удобно переписать поля $\chi$, $\chi^+$ в компонентах:
\begin{equation}
\chi = \left(
\begin{array}{cc}
    0     & \eta \\
    -\eta & 0
\end{array}
\right), \:
\chi^+ = \left(
\begin{array}{cc}
    0     & -\eta^* \\
    \eta^* & 0
\end{array}
\right),
\label{chi_eta}
\end{equation}
где введены комплексные скалярные поля $\eta$, $\eta^*$.
Тогда действие (\ref{action_m_a}) примет вид: 
\begin{gather}
S(\eta, \eta^*,\mathbf{A},m) =
    2 \nabla\eta \nabla\eta^* +
    2 \tau \eta^*\eta +
    \frac{1}{2}(\nabla \times  \mathbf{A})^2 +
    \frac{m^2}{2}  +
    g \eta^*\eta \eta^*\eta \ + \nonumber\\
    +\ 2e^2 \mathbf{A}^2 \eta^*\eta + 4ie \mathbf{A} \eta^* \nabla\eta + 2g_{3} \eta^* \eta m, 
\label{action_eta_m_a}
\end{gather}
где введен новый заряд $g=g_1+g_2/2$. Стоит отметить, что объединение зарядов $g_1$ и $g_2$ в единый заряд $g$ является спецификой случая с двумя спиновыми проекциями. 

\subsection{УФ-ренормировка}
Мы используем размерную регуляризацию $d=4-\varepsilon$, при которой расходимости функций Грина имеют вид полюсов по $\varepsilon$. Устранение расходимостей происходит за счет мультипликативной ренормировки полей и параметров модели. Константы ренормировки вводим следующим образом:
\begin{equation}
        \eta_0=\eta Z_{\eta}, \: 
    m_0=m Z_m, \: 
    g_{0}=g\mu^{\varepsilon}Z_{g}, \: 
    g_{30}=g_3 \mu^{\varepsilon/2}Z_{g_3}, \: 
    \mathbf{A}_0=\mathbf{A} Z_A, \: 
    e_0=e \mu^{\varepsilon/2} Z_e,
\label{renormilized_charges}
\end{equation}
где $\mu$ --- ренормировочная масса, индекс <<0>> обозначает затравочные поля и заряды (именно их значения выражаются через микроскопические параметры фермионной системы, что обсуждалось при выводе действия (\ref{action_0}), однако мы опускали индекс <<0>> до сих пор), $Z_i$ --- соответствующие константы ренормировки.
Будем использовать схему минимальных вычитаний, в которой все константы ренормировки имеют вид
\begin{equation}
       Z_i=1+\sum\limits_{p=1}^{\infty} A_{ip}\left( g,g_3,e \right)\varepsilon^{-p},
\label{min_sub_scheme}
\end{equation}
где $A_{ip}\left( g,g_3,e \right)$ --- некоторая функция зарядов модели. Ренормировочные константы обеспечивают сокращение расходимостей в один-неприводимых функциях Грина:
\begin{gather}
\langle \eta \eta^*\rangle, \: 
\langle mm \rangle, \: 
\langle A_i A_j \rangle, \: 
\langle \eta \eta^*\eta \eta^* \rangle, \:  
\langle \eta A_i \eta^*\rangle,\:
\langle\eta A_i \eta^* A_i\rangle, \: \langle \eta \eta^* m\rangle,   
\end{gather}
где $\langle \cdots \rangle$ обозначает усреднение с весом $\mathrm{exp}(-S)$.
Однопетлевые вычисления дают следующий результат (диаграммная техника и подробности расчета приведены в Приложении): 
\begin{gather}
Z_g = 1+\left(\frac{5}{2}g+12\frac{e^4}{g}-6g_3^2+4\frac{g_3^4}{g}-6e^2\right)\frac{1}{8\pi^2\varepsilon}, \nonumber\\  
Z_{g_3} = 1+\left( g-\frac{3}{2}g_3^2-3e^2 \right)\frac{1}{8\pi^2 \varepsilon}, \label{Z_charges} \\  
Z_e = 1+\frac{e^2}{6}\frac{1}{8\pi^2\varepsilon},\nonumber\\
Z_{\eta} = 1+\frac{3e^2}{2}\frac{1}{8\pi^2\varepsilon}, \quad
Z_m = 1+\frac{g_3^2}{2}\frac{1}{8\pi^2\varepsilon}, \quad
Z_A = Z_e^{-1},\nonumber
\end{gather}
последнее равенство подтверждается тождествами Уорда. РГ-функции модели определяются стандартно \cite{Vasiljev0}:
\begin{align}
\gamma_{\varphi} \equiv \widetilde{\mathcal{D}}_{\mu}\ln{Z_{\varphi}}, \: \:
\beta_u \equiv \widetilde{\mathcal{D}}_{\mu}u,
\label{RG_fuctions_general}
\end{align}
где $\widetilde{\mathcal{D}}_{\mu} = \mu\partial_{\mu}$ при фиксированных затравочных зарядах $u_0$, $u$ принадлежит множеству зарядов $u \in \{g,g_3,e\}$, $\varphi$ --- множеству полей (используется безмассовая схема ренормировки). Из определения получаем следующие $\beta$-функции зарядов:
\begin{gather}
\beta_g = - g \varepsilon + 
g\left( 
    \frac{5}{2} g + 12 \frac{e^4}{g} - 6g_3^2 + 4\frac{g_3^4}{g}-6e^2
\right)\frac{1}{8\pi^2}, \nonumber\\ 
\beta_{g_3} = - g_3 \frac{\varepsilon}{2} +
g_3\left( 
    g - \frac{3}{2} g_3^2 - 3e^2
\right)\frac{1}{8\pi^2}, \label{beta_func} \\  
\beta_e = -e \frac{\varepsilon}{2} + 
\frac{e^3}{6}\frac{1}{8\pi^2}.\nonumber
\end{gather}
%%%%%%%%%%%%%%%%%%%%%%%%%%%%%%%%%%%%%%%%%%%%%%%%%%%%%%%%%
\subsection{Стационарные точки}
Возможные режимы критического поведения в модели определяются асимптотическим поведением инвариантных зарядов $\overline{u}$, подчиняющихся системе линейных дифференциальных уравнений:
\begin{align}
\partial_t \overline{u} \left( t, \{ u \} \right) = \beta_u \left(\{ \overline{u} \}\right), \:\:\:
\overline{u}\left( 0, \{ u \} \right) = u,
\label{ODU}
\end{align}
где $t=\ln{p/\mu}$, $p$ --- импульс. Инфракрасная асимптотика ($t \rightarrow -\infty$) функций Грина определяется набором фиксированных точек $\{u_*\}$ системы (\ref{ODU}), которые даются соотношениями 
\begin{align}
    \beta _u\left( \{u_*\} \right)=0.
\label{fixed_points_general}
\end{align}
Устойчивость фиксированной точки определяется собственными числами матрицы
\begin{align}
     \omega_{kl} = \partial \beta_{u_k} / \partial u_l |_{ \{u\}=\{u_*\} }.
\label{matrix_omega}
\end{align}
Выражения (\ref{beta_func}) определяют два набора фиксированных точек с $e_*=0$ и с $e_* \neq 0$. Последний набор содержит только комплексные значения $g_*$ и $g_{3*}$ и, соответсвенно, не является физически определенным. В таблице \ref{eigen_values} приведены неподвижные точки для набора с $e_*=0$ и отвечающие им собственные числа матрицы $\omega$. Видно, что все эти точки являются седловыми, а не ИК-притягивающими, причем отрицательное собственное число $-\varepsilon/2$ ассоциировано с зарядом $e$. 

\begin{figure}[h!]
\centering
\includegraphics[scale=0.5]{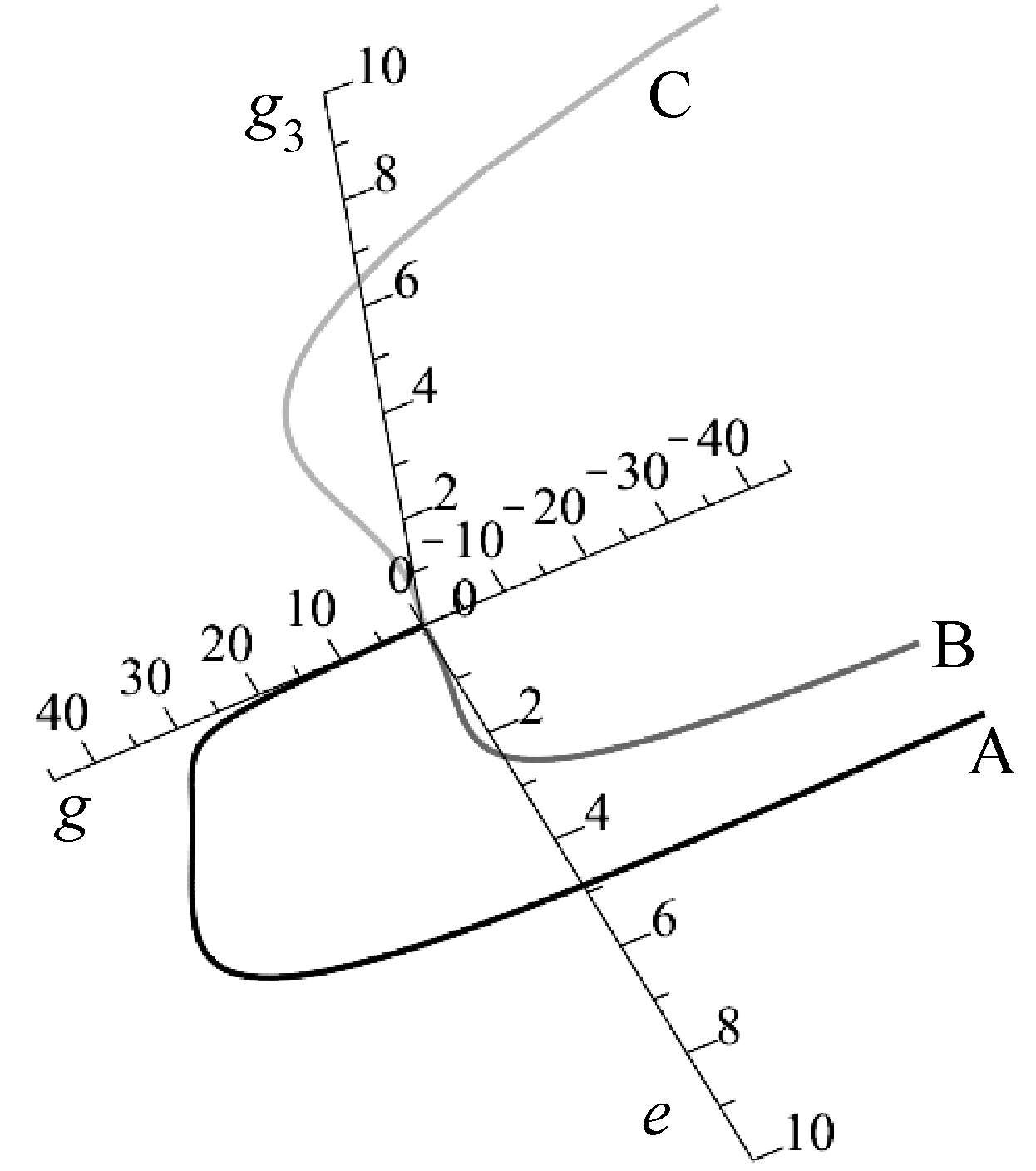}
\caption{Траектории инвариантных зарядов при $\varepsilon=1$ при различных начальных условиях: $A - (0.1,\ 0.01,\ 0.01)$, $B - (0.01,\ 0.1,\ 0.01)$, $C - (0.01,\ 0.01,\ 0.1)$ в координатах $(g,e,g_3)$.} 
\label{trajectories}
\end{figure}

Для понимания поведения системы построим траектории инвариантных зарядов, численно решая систему уравнений (\ref{ODU}). Результат представлен на рис. \ref{trajectories}.
\renewcommand{\arraystretch}{1.2}
\begin{table}[h!]
\centering
\begin{tabular}{|c c | c |} 
 \hline
 $g_*$ & $g_{3_*}$ & \text{Собственные числа} \\
 \hline
 0 & 0 & $-\varepsilon,-\varepsilon/2,-\varepsilon/2$ \\ 
 \hline
 $16/5 \pi^2 \varepsilon$ & 0 & $\varepsilon,-\varepsilon/10,-\varepsilon/2$  \\
 \hline
 $32/5 \pi^2 \varepsilon$  & $2\sqrt{2\varepsilon/5}\pi$ & $\varepsilon/5,\varepsilon, -\varepsilon/2$  \\
 \hline
$16 \pi^2 \varepsilon$ & $2\sqrt{2\varepsilon}\pi$ & $-\varepsilon,\varepsilon,-\varepsilon/2$  \\ 
 \hline
\end{tabular}
 \caption{Неподвижные точки ($e_*=0$) и отвечающие им собственные числа матрицы $\omega$.}
 \label{eigen_values}
\end{table}
Заряд $g$ становится отрицательным в ИК-пределе и покидает область устойчивости (\ref{stability}). Такое же поведение было обнаружено в пятипетлевом приближении для модели (\ref{action_0}) с более чем двумя значениями индекса электронного поля (связанными со старшими спинами, подрешетками или слоистыми структурами материала). На основании анализа поведения свободной энергии было показано, что при этом действительно 
имеет место фазовый переход I рода \cite{Kalagov}, а не слабый фазовый переход I рода ($g \rightarrow +\infty$), который был обнаружен в однопетлевом приближении в моделях с векторным параметром порядка в присутствии калибровочного поля \cite{Dudka_gauge_fields, Halperin_gauge_fields}.

Стоит отметить, что характер фазового перехода в сверхпроводящее состояние до сих пор не известен \cite{Kleinert05}. Действительно, рассмотрим Таблицу~\ref{eigen_values}, исключив из неё собственные числа матрицы $\omega$, относящиеся к магнитному полю. При этом мы обнаруживаем ИК-устойчивую фиксированную точку~№3 с ненулевым значением заряда $g_3$. Однако при $e=0$ рассматриваемая нами модель (\ref{action_eta_m_a}), очевидно, эквивалентна двухкомпонентной статической модели C, для которой показана устойчивость фиксированной точки~№2 \cite{Vasiljev0}. Данный результат, однако, был надежно обоснован лишь с учетом адекватного пересуммирования результатов четырехпетлевых расчетов. 

Аналогичная ситуация, возможно, имеет место для нашей модели (\ref{action_eta_m_a}) при $g_3=0$, т.е. для модели (\ref{action_a}). Хотя однопетлевой расчет и согласуется с нашим утверждением об отсутствии ИК-устойчивых фиксированных точек, результаты двухпетлевых расчетов ставят под сомнение данное утверждение \cite{Folk1996}. К сожалению, в работе \cite{Folk1996} был использован метод Паде-пересуммирования $\varepsilon$ - разложения, который, строго говоря, не применим при описании критического поведения. Заметим, что такая неопределенность, в связи с асимптотическим характером рядов теории возмущений, относится лишь к собственным значениям матрицы~$\omega$, т.е. к анализу устойчивости фиксированных точек. Для числа и координат фиксированных точек результаты однопетлевых расчетов служат достаточно надежным начальным приближением.

\section{Динамика фазового перехода в сверхпроводящее состояние}

На основе статического действия $S(\varphi)$ для набора полей $\varphi \equiv \{\varphi_a\}$ можно построить динамические уравнения, которые в общем случае даются формулой \cite{Vasiljev0} 
\begin{equation} 
     \partial_t \varphi_a = 
     - (\alpha_{ab} + \beta_{ab}) \frac{\delta S(\varphi)}{\delta \varphi_b} + \xi_a, 
     \ 
     \langle \hat{\xi}_a(x)\hat{\xi}_b(x^\prime )\rangle = 
     2\alpha_{ab}\delta\left( x-x^\prime \right),
\label{stoch_equat_general}
\end{equation}
где $\alpha_{ab}$ --- коэффициенты Онзагера, $\beta_{ab}$ --- коэффициенты межмодовой связи, $\xi_a$ --- случайная сила --- белый шум с гауссовым распределением и заданным выше коррелятором, который моделирует все быстроосциллирующие и мелкомасштабные вклады (жесткие моды). Набор полей $\varphi$ должен включать все мягкие моды, существенные для описания гидродинамики или критической динамики рассматриваемой системы, поскольку в обоих упомянутых приближениях малым параметром выступает волновой вектор. Для коэффициентов $\alpha_{ab}$ требуется симметричность и пропорциональность оператору Лапласа или константе в зависимости от того, является ли поле $\varphi_a$ плотностью сохраняющейся величины или нет. Достаточные условия на коэффициенты межмодовой связи приведены в \cite{Vasiljev0}, в частности, $\beta_{ab}$ могут локально зависеть от полей, причем
\begin{align*}
\frac{\delta \beta_{ab}}{\delta \varphi_a}=0
\end{align*}
(суммирование по повторяющемуся индексу подразумевается). Данная форма уравнений обеспечивает их согласованность (при подходящем выборе распределения случайной силы) с достижением равновесного предела при больших временах релаксации. Отметим, что все известные гидродинамические (в широком смысле --- диффузии, теплопроводности, Навье--Стокса) уравнения релаксации на гидродинамической стадии, когда в системе уже сформировались небольшие и плавные отклонения локального среднего значения параметра порядка от равновесного значения, удовлетворяют форме (\ref{stoch_equat_general}) без случайной силы. 

Аналогично общему выражению (\ref{stoch_equat_general}) напишем систему стохастических гидродинамических уравнений для описания динамики в окрестности точки фазового перехода в сверхпроводящее состояние в следующем виде:
\begin{gather}
\begin{cases}
\partial_{t} \chi = 
- \alpha_{\chi} (1+ib)\left[
    - \Delta\chi +
    2ie A_{i} \nabla_{i} \chi +  
    e^2 A^2 \chi + 
    \tau \chi +
    %\right. \\ \left. + 
    \frac{g_1}2\mathrm{Tr}(\chi\chi^{+})\chi +
          \frac{g_2}2 \chi \chi^+ \chi +
    g_3 m \chi
\right] + \\ 
+ ig_4\chi \left[
     m + g_3 \mathrm{Tr} \left( \chi\chi^{+} \right)
\right] + \xi_{\chi\alpha\beta},  
\\
\partial_{t} \chi^+ = 
- \alpha_{\chi} (1-ib)\left[
    - \Delta\chi^+ -
    2ie A_{i} \nabla_{i} \chi^+ +
    e^2 A^2 \chi^+ + 
    \tau \chi^+ +
    %\right. \\ \left. +
    \frac{g_1}2\mathrm{Tr}(\chi\chi^{+})\chi^+ +
          \frac{g_2}2 \chi^+ \chi \chi^+ +
    g_3 m \chi^+
\right] - \\
- ig_4\chi^+\left[
     m + g_3 \mathrm{Tr} \left( \chi\chi^{+} \right)
\right] + \xi_{\chi\alpha\beta}^+, 
\\
\partial_t m =
\alpha_{m} \Delta\left[
     m + g_3 \mathrm{Tr} \left( \chi\chi^{+} \right)
\right] +
ig_4 \mathrm{Tr}\left[
    \chi^+\Delta\chi - \chi\Delta\chi^+ -
2ie A_i \nabla_i(\chi^+\chi)
\right] + \xi_{m}, 
\\
\partial_{t}A_{i} =
\alpha_{A} \left[
    \Delta A_{i} -
    ie\mathrm{Tr}\left(
        \chi^+ \nabla_{i} \chi -\chi \nabla_{i} \chi^+
    \right) -
    2 e^2 A_{i} \mathrm{Tr}\left( \chi^+ \chi \right)
\right] + \xi_{A_i}.
\end{cases}
\label{dynamics_with_m_A}
\end{gather}
Напомним, что используется калибровка Кулона векторного потенциала (в последнем уравнении подразумеваются поперечные проекторы), суммирование по повторяющимся значкам подразумевается, след берется по спиновым значкам.

Поясним данную форму уравнений. Здесь $\alpha _m$ --- коэффициент температуропроводности, который может быть легко измерен экспериментально, в отличие от аналогичных коэффициентов остальных полей. Оператор Лапласа в динамическом уравнении для поля $m$ объяснятся тем, что поле $m$ соответствует сохраняющейся величине. По члену $ig_4\mathrm{Tr}\left(2ieA_i\nabla_i(\chi^+\chi)\right)$ в уравнении на $\partial_t m$ можно заключить, что коэффициент межмодовой связи $g_4$ отвечает, в частности, за изменение температуры за счет переноса плотности куперовских пар магнитным полем. Мнимая единица перед $g_4$ обеспечивает вещественность $\partial_t m$. Коэффициент межмодовой связи $b$ --- стандартный для модели~F динамики критического поведения: он возможен по самой форме построения стохастических или гидродинамических уравнений. В силу изотропности, коэффициент $\alpha_{\chi \alpha \gamma}$ (в общем случае матричный) равен $\alpha_{\chi} \delta_{\alpha \gamma}$.

Следует пояснить, почему отсутствуют члены межмодовой связи между полем $A$ и остальными полями. Для этого приведем выражения
\begin{align*}
\frac{\partial S}{\partial A_i} = 
-\Delta A_i +
ie\mathrm{Tr}(\chi^+\nabla_i\chi - \chi\nabla_i\chi^+) +
2e^2 A_i \mathrm{Tr}(\chi^+\chi)
\end{align*}
и
\begin{align*}
\frac{\partial S}{\partial m}=m+g_3\mathrm{Tr}(\chi^+\chi).
\end{align*}
Симметрийные соображения для внесения в уравнения на $\partial _t \chi$, $\partial _t \chi^+$ межмодовой связи, содержащей ${\partial S}/{\partial A_i}$, требуют появления в коэффициенте либо поля $A_i$, либо оператора $\nabla_i$, что делает такие члены несущественными в критической области и в смысле вывода гидродинамических уравнений. Аналогично, $A_i$ или $\nabla_i$ должен появиться в коэффициенте при ${\partial S}/{\partial A_i}$ в уравнении $\partial _t m$. Однако член пропорциональный $A_i$ запрещен законом сохранения поля $m$. Пропорциональный же градиенту член межмодовой связи дает чисто продольный вклад в уравнение на $\partial _t A$ и запрещен условием калибровки магнитного поля.

Напомним, что уравнения (\ref{dynamics_with_m_A}) (без случайных сил) являются гидродинамическими уравнениями описания динамики систем, способных перейти в сверхпроводящее состояние в случае фазового перехода как второго рода, так и первого. Учет нелинейных членов необходим для описания критической динамики сверхпроводящего фазового перехода (т.е. динамики перехода второго рода). Однако приведенные выше результаты РГ-исследования статического действия (\ref{action_m_a}) не свидетельствуют о непрерывном фазовом переходе, и в таком случае нелинейные члены, скорее всего, несущественны. Мы приводим систему уравнений с учетом нелинейных членов ради того, чтобы провести ниже анализ частных случаев, в которых проявляется фазовый переход второго рода. Более того, так как фазовый переход в сверхпроводящее состояние представляет из себя достаточно интересную и сложную проблему, данный полный вид уравнений может оказаться полезным в дальнейшем для описания тех или иных эффектов.

Итак, в полном объеме использовать систему в настоящий момент не представляется возможным, однако некоторые физические заключения возможны и на данном этапе, поскольку действие (\ref{action_m_a}) содержит в себе части действий других, более исследованных в настоящий момент, моделей. В этом анализе мы ограничимся случаем обычных электронов в качестве носителей заряда ($r=2$), поскольку наличие дополнительных индексов у электронного поля не только сильно усложняет задачу, но и приводит к фазовому переходу первого рода даже в отсутствие магнитного взаимодействия \cite{Kalagov}. При этом из модели можно исключить заряд $g_2$ и использовать статическое действие (\ref{action_eta_m_a}). 

Подчиняясь общим принципам, мы ввели в (\ref{dynamics_with_m_A}) значительное число коэффициентов, экспериментальное определение которых является крайне сложным. Более того, и для статического действия (\ref{action_m_a}) возникает вопрос о существенности тех или иных взаимодействий физических полей. Например, если оказывается возможным пренебречь температурными эффектами, то уравнения динамики получаются из (\ref{dynamics_with_m_A}) тривиально занулением поля $m$ ($g_3=0$).

Исключить из уравнений поле $\mathbf{A}$ можно будет в случае, если многопетлевые вычисления подтвердят результаты приведенного выше РГ-анализа, который показал, что единственными фиксированными точками, претендующими на ИК-устойчивость, являются фиксированные точки с $e_*=0$. Альтернативно, магнитные взаимодействия могут оказаться несущественными при определенных значениях констант взаимодействия, которые сильно зависят от конкретных микроскопических и термодинамических характеристик рассматриваемого сверхпроводника. Тогда наша динамическая система уравнений сводится к модели F критической динамики в стандартной классификации \cite{hohenberg-1977, Vasiljev0}.
В наших переменных она имеет вид 
\begin{eqnarray}
\begin{cases}
\partial_{t} \eta =
- \alpha_{\eta} (1+ib) \left(
    - \Delta\eta + \tau\eta + g|\eta|^2\eta + g_3m\eta
\right) +
ig_4\eta(m + 2g_3|\eta|^2) + \xi_{\eta},
\\
\partial_{t} \eta^* =
- \alpha_{\eta} (1-ib) \left(
    - \Delta\eta^* + \tau\eta^* + g|\eta|^2\eta^* + g_3m\eta^*
\right) -
ig_4\eta^*(m + 2g_3|\eta|^2) + \xi_{\eta^*}, 
\\
\partial_t m = 
\alpha_{m} \left( \Delta m + 2g_3 \Delta |\eta|^2 \right) +
2ig_4(\eta^*\Delta\eta - \eta\Delta\eta^*) + \xi_{m}.
\end{cases}
\label{stoch_equat_case_E}
\end{eqnarray}

По результатам многопетлевых расчетов известно, что в критической области данная модель сводится к модели Е с нулевыми значениями в фиксированной точке коэффициента межмодовой связи $b$ и заряда $g_3$ \cite{Vasiljev0}. Критическое поведение в данной модели до сих пор не исследовано: в ней имеются две фиксированные точки, подозреваемые на ИК-устойчивость, и двухпетлевые расчеты не позволили выбрать единственную правильную. Однако в работах \cite{Juha, Zavoronkov} было показано, что данная модель неустойчива относительно введения звуковых гидродинамических мод (полей скорости и плотности в сжимаемом флюиде) и в критической области сводится к простейшей модели A. Соответственно, в рассматриваемом случае в качестве нарушающих устойчивость мод может фигурировать поле импульса электронной жидкости.

Предположим также аномальную малость или отсутствие коэффициента межмодовой связи $g_4$. При этом рассматриваемая модель оказывается эквивалентна двухкомпонентной модели C критической динамики.
\begin{eqnarray}
\begin{cases}
\partial_{t} \eta =
- \alpha_{\eta} (1+ib) \left(
    - \Delta\eta + \tau\eta + g|\eta|^2\eta + g_3m\eta
\right) + \xi_{\eta},
\\
\partial_{t} \eta^* =
- \alpha_{\eta} (1-ib) \left(
    - \Delta\eta^* + \tau\eta^* + g|\eta|^2\eta^* + g_3m\eta^*
\right) + \xi_{\eta^*}, 
\\
\partial_t m = 
\alpha_{m} \left( \Delta m + 2g_3 \Delta |\eta|^2 \right) + \xi_{m}.
\end{cases}
\label{stoch_equat_case_m}
\end{eqnarray}
Стоит отметить, что в отсутствии магнитного поля у статического действия 
\begin{gather}
S(\eta, \eta^*,m) =
    2 \nabla\eta \nabla\eta^* +
    2 \tau \eta^*\eta  +
    \frac{m^2}{2}  +
    g \eta^*\eta \eta^*\eta  + 2g_{3} \eta^* \eta m 
    \nonumber
\end{gather}
появляется ИК-устойчивая точка ($g_* =32/5 \pi^2 \varepsilon$; $g_{3_*}=2\sqrt{2\varepsilon/5}\pi$), см. Таблицу \ref{eigen_values}. Однако 
на основании трехпетлевых расчетов в данной модели также было показано, что в двухкомпонентном случае ИК-устойчивой точкой является точка ($g_*=16/5 \pi^2 \varepsilon$; $g_{3_*}=0$) \cite{Vasiljev0}. В критической области параметр $g_3$ стремится к нулю и модель превращается в модель~A.

В случае, когда по каким-либо причинам в критической области оказываются несущественными и магнитное поле, и поле температуры мы немедленно получаем динамические уравнения в виде: 
\begin{eqnarray}
\begin{cases}
\partial_{t} \eta =
- \alpha_{\eta} (1+ib) \left(
    - \Delta\eta + \tau\eta + g|\eta|^2\eta 
\right) + \xi_{\eta},
\\
\partial_{t} \eta^* =
- \alpha_{\eta} (1-ib) \left(
    - \Delta\eta^* + \tau\eta^* + g|\eta|^2\eta^* 
\right) + \xi_{\eta^*}
\end{cases}
\label{stoch_equat_case_chi}
\end{eqnarray}
Согласно \cite{Vasiljev0}, это двухкомпонентная A-модель, свойства которой хорошо известны. Динамический критический индекс $z$ вычислен в работе \cite{nalimov-2009} и равен $z=2.014^{+0.011}_{-0.0}$ (4-петлевой расчет). На основании 5-петлевых расчетов диаграмм в работе \cite{Adzhemyan_2022} было получено значение $z=2.0246(10)$.

\section{Заключение}
Одним из выводов нашей работы следует признать демонстрацию сложности введения корректного динамического уравнения релаксации в критической области исходя из чисто феноменологических соображений.
Модель, содержащая лишь поле параметра порядка, показывает фазовый переход II рода, и динамика в таком случае сводится к модели~А с известным динамическим критическим индексом. Учет поля температуры оказывается несущественным: в критической точке межмодовая связь исчезает и динамика параметра порядка сводится к той же модели~А. Однако дополнительный учет магнитного взаимодействия радикально меняет картину: устойчивая неподвижная точка становится седловой, система претерпевает фазовый переход I рода. Динамика такой модели описывается уравнениями (\ref{dynamics_with_m_A}), где уже и температурное, и магнитное поля становятся существенными. Стоит отметить, что вопрос о характере фазового перехода в сверхпроводящее состояние до сих пор является открытым \cite{Kleinert05} и данная работа показывает, что введение в модель дополнительных мод может сильно изменить картину фазового перехода. Благодаря подходу \cite{Komarova_2013} удается показать, по крайней мере в однопетлевом приближении, что учет магнитного поля приводит именно к фазовому переходу I рода, а не к слабому переходу I рода, как получалось при использовании модели с векторным параметром порядка \cite{Dudka_gauge_fields}.

Мы получили стохастические уравнения динамики для бозонных критических мод (\ref{dynamics_with_m_A}), развивая предложенный в \cite{Komarova_2013} подход к описанию сверхпроводимости. В простейшем случае ($r=2$, $g_3=0$) полученные нами уравнения совпадают по форме с уравнениями Гинзбурга--Ландау, что не удивительно, поскольку мы работаем в рамках приближения линейной гидродинамики. Отличие, во-первых, заключается в матричном характере параметра порядка. Во-вторых, ренормгрупповой анализ модели позволяет корректно описывать критическую окрестность, давая эффективные значения зарядов в критической точке, затравочные значения которых вычисляются через микроскопические характеристики системы фермионов. Во-третьих, подход к построению стохастических уравнений (\ref{stoch_equat_general}) обеспечивает возможность достижения равновесия, тем самым позволяя корректно и в единой манере вводить дополнительные поля, такие как температурное поле $m$, и анализировать их существенность в критической окрестности.

Мы также привели соображения, свидетельствующие о том, что при непрерывном фазовом переходе в сверхпроводящее состояние критическая динамика описывается достаточно точно вычисленным в настоящее время индексом $z$ стохастической модели A.

Наша работа свидетельствует о том, что необходимы по крайней мере трехпетлевые вычисления в статической модели (\ref{action_m_a}) с последующим адекватным пересуммированием результатов для определения типа фазового перехода в сверхпроводящее состояние. Также может оказаться необходимым использование последовательно микроскопического подхода к описанию динамики рассматриваемого фазового перехода на основе формализма временных функций Грина при конечной температуре для выяснения существенных для описания критической динамики <<мягких>> полей. 

\section{Благодарности}

Исследование Л.А.~Гостевой поддержано Министерством науки и высшего образования Российской Федерации (соглашение № 075–15–2022–287 от 06.04.2022).

\section{Приложение}
В данном разделе приведена диаграммная техника, которая использовалась при вычислении констант ренормировки для модели с магнитным взаимодействием и температурным полем (\ref{action_eta_m_a}). 
В Таблице \ref{propagator} введены обозначения для пропагаторов рассматриваемой модели. Крестик на линии обозначает комплексное сопряжение, а $\langle \cdots \rangle_0$ обозначает нулевое приближение в ряду для соответствующей функции Грина. Так же в Таблице \ref{propagator} указано импульсное представление каждого пропагатора. Вершины модели представлены в Таблице \ref{vortex}.

В Таблице \ref{all_diagrams} приведены все диграммы в нулевом и в однопетлевом приближениях в ряду для соответсвующих один-неприводимых функций Грина, которые требуют устранения расходимостей. Под петлевыми диаграммами указан результат вычисления расходимости соответствующего петлевого подграфа в безмассовой схеме ренормировки при размерной регуляризации $d=4-\varepsilon$.

\begin{table}[h!]
    \centering
    \begin{tabular}{l}
           $\langle \eta \eta^* \rangle_0$ \ = \ \includegraphics[]{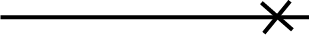} \ = \ $1/(2p^2)$ \\
           $\langle m m \rangle_0$ \ = \ \includegraphics[]{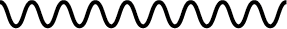} \ = \ $1/2$ \\
           $\langle A_i A_j \rangle_0$ \ = \ \includegraphics[]{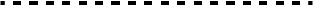} \ = \ $(\delta_{ij}-p_i p_j/p^2)/p^2$
    \end{tabular}
    \caption{Пропагаторы теории.}
    \label{propagator}
\end{table}

\begin{table}[h!]
\centering
\begin{tabular}{llll}
\includegraphics[scale=1]{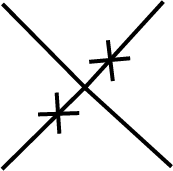} & \includegraphics[scale=1]{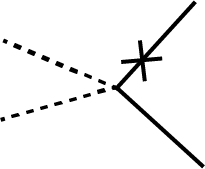} & \includegraphics[scale=1]{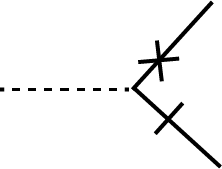} & \includegraphics[scale=1]{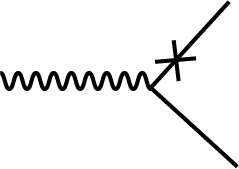} \\
$g \eta^* \eta \eta^* \eta$ &  $2e^2 \mathbf{A}^2 \eta^* \eta$ & $4ie \mathbf{A} \eta^* \nabla \eta$ & $2 g_{3} \eta^* \eta m$
\end{tabular} 
\caption{Вершины теории. Палочка на линии обозначает импульс $ip_i$, возникающий от производной в члене $4ie \mathbf{A} \eta^* \nabla \eta$.}
\label{vortex}
\end{table} 

\renewcommand{\arraystretch}{2}
\begin{table}[h!]
\centering
\begin{tabular}{|c|c|c|c c c c}%{|m|m|m|m m m m}
\cline{1-6}
\multirow{2}{*}{$\langle \eta \eta^* \eta \eta^* \rangle$} &
\includegraphics[scale=0.4]{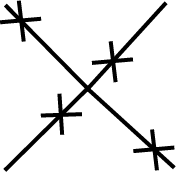} &
\includegraphics[scale=0.4]{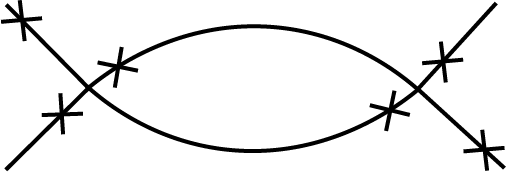} &
\multicolumn{1}{c|}{\includegraphics[scale=0.4]{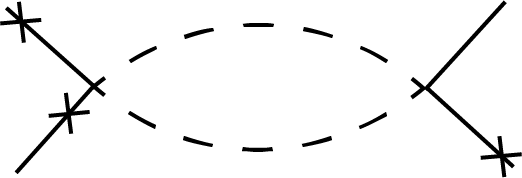}} & 
\multicolumn{1}{c|}{\includegraphics[scale=0.4]{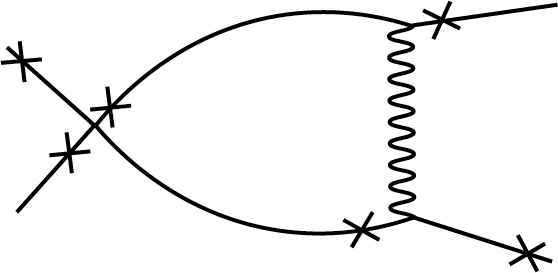}} &
\multicolumn{1}{c|}{\includegraphics[scale=0.4]{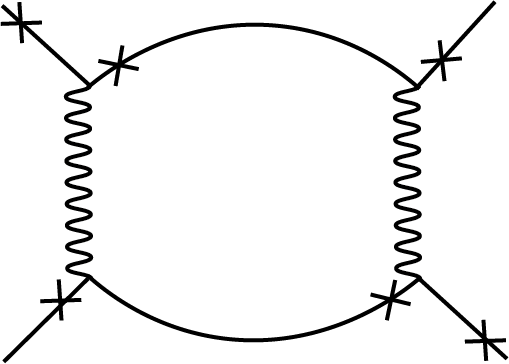}} & \\
\cline{2-6}
& 
$-4g$ & 
$ \frac{5 g^2}{4 \pi^2\varepsilon}$ & 
\multicolumn{1}{c|}{$\frac{6e^4}{\pi^2\varepsilon}$} & 
\multicolumn{1}{c|}{$-\frac{3gg_3^2}{\pi^2 \varepsilon}$} & 
\multicolumn{1}{c|}{$\frac{2 g_3^4}{\pi^2 \varepsilon}$}  & \\
\cline{1-6}
%%%%%%%%%%%%%%%%%%%%%%%%%%%%%%%%%%%%%%%%%%%%%%%%%%%%%%%%%%%%%%%%%%%%%
\multirow{2}{*}{$\langle \eta \eta^* \rangle$} & 
\includegraphics[scale=0.4]{eta_1.png} & 
\centering{\includegraphics[scale=0.55]{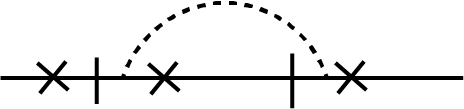}} & & & & \\ 
\cline{2-3}
& 
$1/2$ &  
$\frac{3e^2}{4\pi^2 \varepsilon}$ & & & &  \\
\cline{1-3}
%%%%%%%%%%%%%%%%%%%%%%%%%%%%%%%%%%%%%%%%%%%%%%%%%%%%%%%%%%%%%%%%%%%%%
\multirow{2}{*}{$\langle A_i A_j \rangle$} & 
\includegraphics[scale=0.4]{A_1.png} &
\includegraphics[scale=0.4]{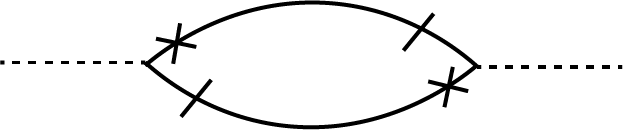} & & & & \\
\cline{2-3}
& 
1 & 
$-\frac{e^2}{24 \pi^2 \varepsilon}$ & & & & \\
\hline
%%%%%%%%%%%%%%%%%%%%%%%%%%%%%%%%%%%%%%%%%%%%%%%%%%%%%%%%%%%%%%%%%%%%%
\multirow{2}{*}{$\langle \mathbf{A} \mathbf{A} \eta \eta^* \rangle$} &
\includegraphics[scale=0.4]{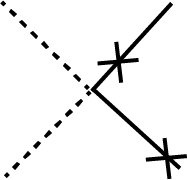} & 
\includegraphics[scale=0.4]{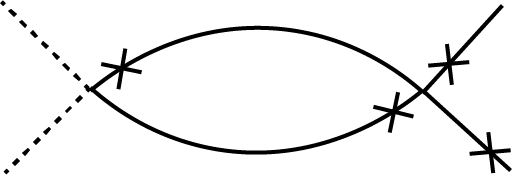} & 
\multicolumn{1}{l|}{\includegraphics[scale=0.4]{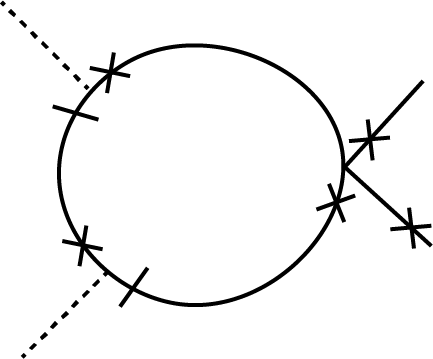}} &
\multicolumn{1}{c|}{\includegraphics[scale=0.4]{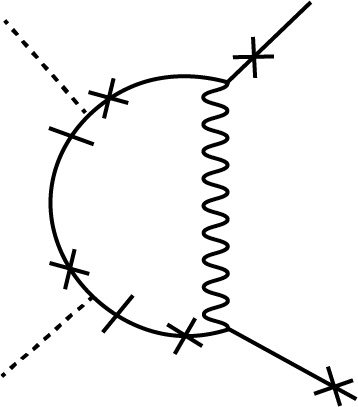}} &
\multicolumn{1}{l|}{\includegraphics[scale=0.4]{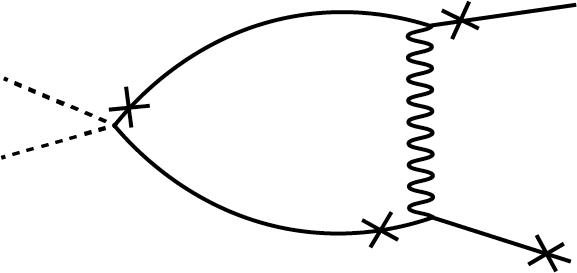}} &
\multicolumn{1}{c|}{\includegraphics[scale=0.4]{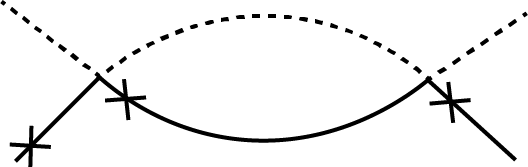}} \\
\cline{2-7} 
& 
$-4e^2$ &
$\frac{ge^2}{2 \pi^2 \varepsilon}$ & 
\multicolumn{1}{c|}{$-\frac{ge^2}{2 \pi^2 \varepsilon}$} & 
\multicolumn{1}{c|}{$-\frac{e^2g_3^2}{2 \pi^2 \varepsilon}$} & 
\multicolumn{1}{c|}{$\frac{e^2g_3^2}{2 \pi^2 \varepsilon}$} & 
\multicolumn{1}{c|}{$\frac{3e^4}{2 \pi^2 \varepsilon}$} \\
\hline
%%%%%%%%%%%%%%%%%%%%%%%%%%%%%%%%%%%%%%%%%%%%%%%%%%%%%%%%%%%%%%%%%%%%%
\multirow{2}{*}{$\langle A_i \eta \eta^* \rangle$} & 
\includegraphics[scale=0.4]{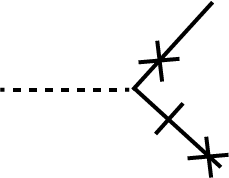} & 
\includegraphics[scale=0.4]{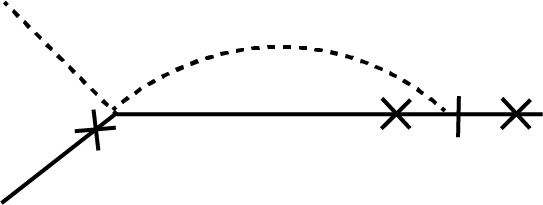} & 
\multicolumn{1}{c|}{\includegraphics[scale=0.4]{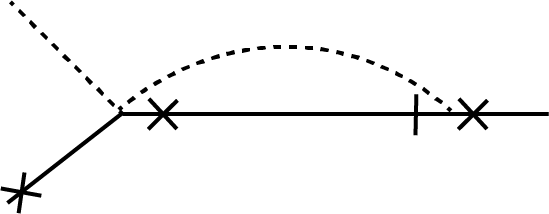}} & & & \\ 
\cline{2-4}
&
$4ie$ &
$-\frac{3ie^3}{4\pi^2\varepsilon}$ & 
\multicolumn{1}{c|}{$-\frac{3ie^3}{4\pi^2\varepsilon}$} & & & \\ 
\cline{1-4}
%%%%%%%%%%%%%%%%%%%%%%%%%%%%%%%%%%%%%%%%%%%%%%%%%%%%%%%%%%%%%%%%%%%%%
\multirow{2}{*}{$\langle m m \rangle$} & 
\includegraphics[scale=0.4]{m_1.png} &
\includegraphics[scale=0.4]{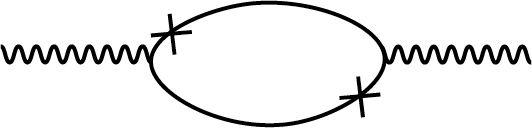} & & & & \\ 
\cline{2-3}
& 
1 &
$\frac{g_3^3}{8\pi^2 \varepsilon}$ & & & & \\ 
\cline{1-4}
%%%%%%%%%%%%%%%%%%%%%%%%%%%%%%%%%%%%%%%%%%%%%%%%%%%%%%%%%%%%%%%%%%%%%
\multirow{2}{*}{$\langle m \eta \eta^* \rangle$} &
\includegraphics[scale=0.4]{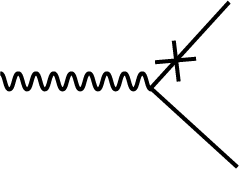} &
\includegraphics[scale=0.4]{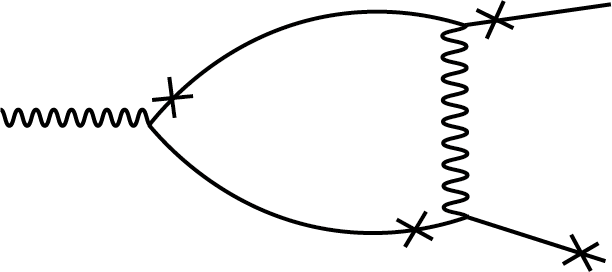} &
\multicolumn{1}{c|}{\includegraphics[scale=0.4]{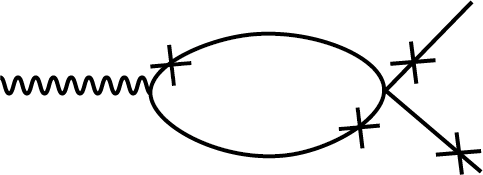}} & & & \\ 
\cline{2-4}
& 
$-2g_3$ &
$-\frac{g_3^3}{4\pi^2 \varepsilon}$ &
\multicolumn{1}{c|}{ $\frac{g g_3}{4\pi^2 \varepsilon}$} & & & \\ 
\cline{1-4}
\end{tabular}
\caption{Диаграммы в нулевом и однопетлевом приближениях.}
\label{all_diagrams}
\end{table}

\newpage
\bibliographystyle{my.bst}
\bibliography{bibliography}

\end{document}